\documentclass[prb,aps,epsf,twocolumn,showpacs,floatfix,superscriptaddress]{revtex4}
\usepackage{graphicx}
\usepackage{epsfig}
\usepackage{latexsym}
\usepackage{amsmath}
\begin{document}
\newcommand{\beq}{\begin{equation}}
\newcommand{\beqr}{\begin{eqnarray}}
\newcommand{\eeqr}{\end{eqnarray}}
\newcommand{\eeq}{\end{equation}}
\newcommand{\s}{{\sigma}}
\newcommand{\e}{{\varepsilon}}
\newcommand{\bofp}{{\bB (\bp )}}
\newcommand{\om}{{\omega}}
\newcommand{\baro}{{\bar{\rho}_{}(\bq )}}
\newcommand{\barop}{{\bar{\rho}_{p}(\bq )}}
\newcommand{\barchi}{{\bar{\chi}_{}(\bq )}}
\newcommand{\Om}{{\Omega}}
\newcommand{\D}{{\Delta}}
\newcommand{\half}{{{1 \over 2}}}
\newcommand{\de}{{\delta}}
\newcommand{\al}{{\alpha}}
\newcommand{\nh}{{\nu = \half}}
\newcommand{\nnh}{{\nu \ne \half}}
\newcommand{\ga}{{\gamma}}
\newcommand{\La}{{\Lambda}}
\newcommand{\be}{{\beta}}
\newcommand{\psib}{{\bar{\psi}}}
\newcommand{\rb}{{\bar{\rho}}}
\newcommand{\phib}{{\bar{\phi}}}
\newcommand{\dt}{{\Delta}}
\newcommand{\dva}{{\frac{\vp\times\va}{2\pi}-\rb}}
\newcommand{\dvab}{{\frac{\vp\times(\va-\vb)}{4p\pi}}}
\newcommand{\w}{{\omega}}
\newcommand{\zh}{{\hat{z}}}
\newcommand{\qh}{{\hat{q}}}
\newcommand{\vA}{{\vec{A}}}
\newcommand{\va}{{\vec{a}}}
\newcommand{\vrr}{{\vec{r}}}
\newcommand{\vj}{{\vec{j}}}
\newcommand{\vE}{{\vec{E}}}
\newcommand{\vB}{{\vec{B}}}
\def\bA{{\mathbf {\cal A}}}
\def\bm{{\mathbf m}}
\def\bsig{{\mathbf \sigma}}
\newcommand{\etab}{\mbox{\boldmath $\eta $}}
\def\bB{{\mathbf {\cal B}}}
\def\bp{{\mathbf p}}
\def\be{{\mathbf e}}
\def\bI{{\mathbf I}}
\def\bn{{\mathbf n}}
\def\bM{{\mathbf M}}
\def\bG{{\mathbf G}}
\def\bq{{\mathbf q}}
\def\bp{{\mathbf p}}
\def\br{{\mathbf r}}
\def\bx{{\mathbf x}}
\def\bs{{\mathbf s}}
\def\bS{{\mathbf S}}
\def\bQ{{\mathbf Q}}
\def\bq{{\mathbf q}}
\def\bs{{\mathbf s}}
\def\bi{{\mathbf i}}
\def\bj{{\mathbf j}}
\def\bG{{\mathbf G}}
\def\bl{{\mathbf l}}
\def\bPi{{\mathbf \Pi}}
\def\bJ{{\mathbf J}}
\def\bR{{\mathbf R}}
\def\bz{{\mathbf z}}
\def\ba{{\mathbf a}}
\def\bk{{\mathbf k}}
\def\bK{{\mathbf K}}
\def\bP{{\mathbf P}}
\def\bg{{\mathbf g}}
\def\bX{{\mathbf X}}
\newcommand{\Psib}{\mbox{\boldmath $\Psi $}}
\newcommand{\thetab}{\mbox{\boldmath $\theta $}}
\newcommand{\sigmab}{\mbox{\boldmath $\sigma $}}
\newcommand{\nablab}{\mbox{\boldmath $\nabla $}}
\newcommand{\gammab}{\mbox{\boldmath $\gamma $}}
\newcommand{\vx}{{\vec{x}}}
\newcommand{\vq}{{\vec{q}}}
\newcommand{\vQ}{{\vec{Q}}}
\newcommand{\vd}{{\vec{d}}}
\newcommand{\vb}{{\vec{b}}}
\newcommand{\vp}{{\vec{\partial}}}
\newcommand{\p}{{\partial}}
\newcommand{\gr}{{\nabla}}
\newcommand{\ra}{{\rightarrow}}
\def\dd{d^{\dagger}}
\def\half{{1\over2}}
\def\fifth{{1\over5}}
\def\third{{1\over3}}
\def\twof{{2\over5}}
\def\threes{{3\over7}}
\def\rhob{{\bar \rho}}
\def\ua{\uparrow}
\def\da{\downarrow}
\def\eqa{\begin{eqnarray}}
\def\eea{\end{eqnarray}}
\def\jetp{{\it Sov. Phys. JETP\ }}
\parindent=4mm
\addtolength{\textheight}{0.9truecm}
\title{The $\nu =\half$ Landau  level: half-full or half-empty?
}
\author{Ganpathy Murthy }
\affiliation{  Department of Physics and
Astronomy, University of Kentucky, Lexington KY 40506-0055   }
\author{R. Shankar}
\affiliation{Department of Physics, Yale University, New Haven CT 06520}

\date{\today}
\begin{abstract}

 We show here that an extension of the Hamiltonian theory developed by
 us over the years furnishes a composite fermion (CF) description of
 the $\nu =\half$ state that is particle-hole (PH) symmetric, has a
 charge density that obeys the magnetic translation algebra of the
 lowest Landau level (LLL), and exhibits cherished ideas from highly
 successful wave functions, such as a neutral quasi-particle with a
 certain dipole moment related to its momentum. We also a provide an
 extension away from $\nh$ which has the features from $\nh$ and
 implements the the PH transformation on the LLL as an anti-unitary
 operator ${\cal T}$ with ${\cal T}^2=-1$.  This extension of our past
 work was inspired by Son, who showed that the CF may be viewed as a
 Dirac fermion on which the particle-hole transformation of LLL
 electrons is realized as time-reversal, and Wang and Senthil who
 provided a very attractive interpretation of the CF as the bound
 state of a semion and anti-semion of charge $\pm {e\over 2}$. Along
 the way we also found a representation with all the features listed above
 except that now ${\cal T}^2=+1$. We suspect it corresponds to an
 emergent charge-conjugation symmetry of the $\nu =1$ boson problem
 analyzed by Read.
\end{abstract}

\maketitle

\section{Introduction to the  problem}

Consider electrons in high magnetic field
\cite{QH-reviews,FQH-reviews} partially filling the Lowest Landau
Level (LLL) in the limit when the cyclotron frequency $\omega_c \to
\infty$ is much larger than the interaction. In this limit one expects
a complete description entirely in terms of the LLL states.  A
partially occupied band of electrons may be equally well described in
terms of electrons on top of an empty band or holes depopulating the
filled band. At filling factor $\nu = \half$, for translationally
invariant two-body interactions, the Hamiltonian has particle-hole
(PH) {\em symmetry} \cite{PH-LKGK}, and one should be able to develop
a treatment in which this symmetry is manifest.  In addition it must
be possible to relate the physics at $\nu$ to that at $1-\nu$ away
from $\nu =\half$ by a PH {\em transformation}.  This had proven
elusive until recent
work\cite{Son,Wang-Senthil,Max-Ashvin,jain2015,gang-of-six}, most notably that
of Son\cite{Son}. It should be noted that this problem is intimately
related to the surface states of strongly correlated,
three-dimensional time-reversal invariant topological
insulators\cite{Wang-Senthil,Max-Ashvin}, and that numerical
work\cite{jain2015,gang-of-six} confirms the particle-hole symmetry of the
$\nh$ ground state in the LLL.

A class of successful approaches to the Fractional Quantum Hall Effect
(FQHE) requires flux attachment, that is, transforming the electrons
into either Composite Bosons (CB)\cite{ZHK} by attaching an odd number
of flux quanta, or Composite Fermions\cite{Jain,Lopez-Fradkin}, by
attaching an even number.  At half-filling this turns the electron
problem to that of CF's that see zero field on average and form a
Fermi liquid, as elucidated by Halperin, Lee, and Read
(HLR)\cite{HLR}. A similar Fermi surface arises in the problem of the
hard-core bosons at $\nu =1$ in the LLL analyzed by
Read\cite{Read-nu=1}. By attaching one flux quantum to each of the
bosons he turns them into fermions that see zero field on the
average.

When implemented in the wave function language and projected to the
LLL, Composite Fermions (CFs) produce excellent parameter-free wave
functions\cite{Jain,FQH-reviews} for the Jain fractions of the type
$\nu=\frac{p}{2p+1}$, and the Rezayi-Read\cite{Rezayi-Read} wave
function for the gapless $\nu =\half$ state.

There are at least two types of particles called CF's in the past
fractional quantum Hall (FQH) literature.  If we work in the complete
Hilbert space of the electron, flux attachment in Chern-Simons (CS)
theory\cite{ZHK,Lopez-Fradkin,HLR} leads to a particle of charge $e$
(the electron charge). For such a particle one can derive, at $\nh$,
the constraint $\sigma_{xy}^{CF}= -\half {e^2 \over h} $ following
Lee, Krotov, Gan, and Kivelson\cite{PH-LKGK}. We shall refer to this
as the Chern-Simons CF. The other is the CF that resides entirely in
the LLL and the one we will focus on.  At $\nh$ attachment of the
double-vortex (now double-zero and not just a phase $4\pi$) drives
away a charge $-e$ and leaves us with a neutral CF. This point of view
was emphasized by Read\cite{Rezayi-Read} who also argued that the CF
of momentum $\bp$ will have a dipole moment $\hat{z} \times \bp l^2$
where

\beq
l^2= {1 \over eB}. 
\eeq

At the moment, there are conflicting claims about $\nu=\half$. On the
one hand are arguments from Chern-Simons theory that there are two
distinct states, a particle-CF-Fermi sea and a hole-CF-Fermi
sea\cite{barkeshli}. On the other hand are numerical calculations in
the wave-function language\cite{jain2015} or by exact
diagonalization\cite{gang-of-six} which show a unique, particle-hole symmetric state at $\nh$.

In this work we will not attempt to resolve the controversy, but begin
with the premise that the $\nu=\half$ state does have particle-hole
symmetry in the lowest Landau level.
 
Now for the main business of this paper. The work of HLR\cite{HLR}
which leads to a Fermi-surface for the CS-CFs is not particle-hole
symmetric at $\nh$. One does not expect it to be since the symmetry is
emergent only in the limit $\omega_0 \to \infty$ or $m_e \to 0$ which
is problematic in their approach. But is there a description in which
PH symmetry is manifest at $\nh$? Is there a way to relate the physics
at $\nu$ to that at $1-\nu$? An affirmative answer was given recently
by Son\cite{Son}. The underlying physical picture was provided by Wang
and Senthil (WS)\cite{Wang-Senthil}. Connections of this problem to
the surface states of 3D time-reversal invariant topological
insulators have also been elucidated in recent
work\cite{Wang-Senthil,Max-Ashvin}.

In this paper we will re-examine and extend our Hamiltonian theory
\cite{MS-Ham} in light of these developments.
 
The heart of our Hamiltonian approach\cite{MS-Ham} is to define the
LLL problem algebraically in terms of the commutation rules of the
projected electron density $\baro$, which alone enters the LLL
Hamiltonian and obeys the magnetic translation algebra.  Having
defined it thus, the next step is to represent this algebra faithfully
in a larger fermionic space subject to some constraints.  The reason
is that in this new space there is a natural Hartree-Fock state at the
Jain fractions and the limiting case $\nu =\half$.

When we re-examine our approach in the light of Son's work\cite{Son}
we find several new results which we report here.

The first is that the most straightforward representation of a neutral
CF in terms of a nonrelativistic one-component CF yields a description
of an LLL system in which there is an anti-unitary operator ${\cal T}$
with ${\cal T}^2=+1$ which plays the role of time-reversal on the CF's
and charge conjugation on the physical charge. It exchanges the role
of a (hard core) boson and a single vortex. Since we know that the
electronic LLL problem must have ${\cal T}^2=-1$\cite{levin-son}
(confirmed by recent numerics\cite{gang-of-six}), this most probably
describes an emergent charge-conjugation symmetry of the $\nu =1$
boson problem in the LLL\cite{Read-nu=1}.

Secondly, after the work of Son\cite{Son} and WS\cite{Wang-Senthil},
we realized that we could represent the magnetic translation algebra
in the space of a two-component Dirac CF, whose number is always half
the number of flux quanta, regardless of the number of electrons. Now
we find exactly what Son did: An anti-unitary operator ${\cal T}$ with
${\cal T}^2=-1$ which plays the role of time-reversal on the CF's and
charge conjugation on the electronic charge.  In addition, we obtain
two representations of the physical charge density at all $\nu$.  One
has a Hamiltonian with a set of constraints commuting with it, but
does not manifestly show the neutrality of the Dirac CF or the PH
symmetry. The other shows both symmetries in a manifest way, but
ignores the constraints that limit the larger space to the LLL sector.

But for Son's work it would not have occurred to us to bring in Dirac
fermions, because we were always insistent on working in a space that
was adiabatically connected to the single-component electron. This was
to ensure that we did not represent the problem algebraically in a
space that had no bearing on the LLL problem. Furthermore, it would
also not have occurred to us, who tried to implement Jain's
construction using operators, to tie the number of CFs to the flux as
Son did (and not the number of electrons). We thank Senthil for
emphasizing the importance of this point, which in the end was what
made it possible to extend the GMP algebra away from $\nu = \half$
when working with Dirac fermions. 

It is our hope that the explicit representation of the electronic
charge density obeying the GMP algebra, and a neutral fermion of the
right dipole moment paves the way for many operator based
calculations.

Now for the organization of the paper. First we will furnish a
telegraphic introduction to the Hamiltonian theory\cite{MS-Ham} citing
only those results germane to this paper.  Next we will consider the
most natural representation of the algebraic problem: in terms of a
one-component non-relativistic CF, whose number is equal to the number
of electrons. This gives us a theory which has a PH symmetry with
${\cal T}^2=1$, possibly pertaining to an emergent charge-conjugation
symmetry of the $\nu=1$ hard-core boson problem\cite{Read-nu=1}. Next
we will demonstrate that the magnetic translation algebra can be
realized in the space of a Dirac CF, whose number is half the number
of flux quanta and ${\cal T}^2=-1$. This representation works both at
$\nu=\half$ and away from it. At  $\nh$ it provides a
representation of the physical charge density that realizes the PH
symmetry in a manifest way. Away from $\nh$, we show how the PH
transformation of the LLL electrons (implemented by ${\cal T}$)
relates $\nu \leftrightarrow 1-\nu$. A summary follows.

\section{ Hamiltonian theory: why and how?}

The problem of interacting electrons in the LLL is defined by the
LLL-projected Hamiltonian

\beq
 \bar{H}= \half \sum_{\bq} \baro v_{ee}(\bq ) \bar{\rho}_{}(-\bq),\label{lllh}
\eeq
where $\baro$ is the electron density projected to the LLL
\beq
\baro = \sum_j  e^{-i \bq \cdot \bR^{e}_{j}}
\eeq
where a factor $e^{-q^2l^2/4}$ from  each $\baro$ has been absorbed in the electron-electron potential $v_{ee}$, and   
 $\bR^e$ is the electron guiding-center coordinate obeying 
\beq
\left[ \bR^{e}_{x}, \bR^{e}_{y}\right] = -il^2.\label{recomm}
\eeq

As a result of Eqn.\ref{recomm}, $\baro $ obeys the Girvin, MacDonald and Platzman\cite{GMP} (GMP) or magnetic translation algebra:

\beq \left[ \bar{\rho}_{}(\bq ),
  \bar{\rho}_{}(\bq' )\right]= 2 i \sin \left[ {
    l^2\over 2 }{\bq \times \bq' }\right] \bar{\rho}_{}(\bq +\bq') .\label{gmp-alg} \eeq 

The mathematical problem is defined by the Hamiltonian $\bar{H}$ and
the GMP algebra of the projected charge density entering it. Of
course, the answer could vary with the space in which we represent
this algebra. (Compare the spin-$\half$ and spin-$1$ chains.)  The original electron problem in defined in the
electronic LLL Hilbert space. Now the trouble with formulating the problem in  the
electronic space is that there is no Hartree-Fock (HF) state due
to the huge degeneracy of the partially filled LLL.

 Jain \cite{Jain}  beats this by switching to the CF which sees a
weaker field and fills exactly $p$ Landau levels for the fractions

\beq
 \nu = { p \over 2p+1}.  
\eeq

Motivated by Jain, we will start by using the Hilbert space of a
one-component non-relativistic fermion that sees just this field to represent the Hamiltonian of
Eq. (\ref{lllh}).  {\em The number of CFs is equal to
the number of electrons in this construction.} 

Let us trace some of the steps along the way to the final picture.
More details are provided in the Appendix.  Starting with the CS
theory of fermions in the full Hilbert space\cite{MS-Ham}, we trade
the CS gauge field (whose components are conjugate) for magnetoplasmon
oscillators \`{a} la Bohm-Pines\cite{Bohm-Pines}.  When we decouple
the oscillators and freeze them in the ground state, we obtain an LLL
description of the projected electron density $\baro$ and the
constraints $\barchi$ that pay for the collective oscillator degrees
of freedom. We could only derive expressions for $\baro$and $\barchi$
at small $q$.  The series were exponentiated \cite{MS-Ham} to form expressions valid
all $q$, with the nice feature that $\baro$ obeyed the
GMP algebra and $\barchi$ commuted with $\baro$ and hence $\bar{H}$
and closed to form a GMP-like algebra of a particle with the charge of
the double-vortex.  

In this connection, which evolves from the CS theory, the number of CFs equals the number of electrons ($N_{CF}= N_e$)
and the (spin-polarized) fermion has only one-component. 

Here is the final picture: The CF  experiences a
weaker magnetic field $B^*$, just right to fill $p$ Landau levels, as
envisaged by Jain.  That is encoded in the CF velocity operator which obeys

\beq
\left[ \bPi^{*}_{x}, \bPi^{*}_{y} \right]= {i eB^*}=ie^*B={i\over l^{*2}}=i{1-c^2\over l^2}\label{Picom}
\eeq

where 
\beq
c^2 = 2\nu = {2p \over p+1}.
\eeq

For example, at $\nu = {1 \over 3}$ we have $p=1, c^2 = {2 \over 3},
e^*= {e \over 3}$ and at $\nu = \half$, we have $c=1$ and $p =
\infty$.

Now we introduce in this (full fermionic) space a pair of coordinates
\beq
 \bR^e = \br -{l^2 \over 1+c}\hat{\bz}\times \bPi^*\label{Re}
\eeq
which obey 
\beq
 \left[ \bR^{e}_{x}, \bR^{e}_{y}\right] = -il^2. \eeq

We recognize this as the algebra of the guiding center coordinate
of the electron.  This ensures that the corresponding density

\beq
   \bar{\rho}_{}(\bq )=\sum_je^{-i \bq \cdot \bR^{e}_{j}}\label {expRe}
\eeq

obeys the GMP algebra.  The LLL-projected Hamiltonian is represented
{\em in the CF space} by

\beqr
 \bar{H}&=& \half \sum_{\bq} \baro v_{ee}(\bq ) \bar{\rho}_{}(-\bq).\label{projH}
 \eeqr
 
{\em Although $\bar{H}$ is now written in terms of a $\baro$ which obeys
the same GMP algebra as the one in Eqn. \ref{lllh}, there is a big
difference.} It is now expressed in terms of CF coordinates in their
Hilbert space, and {\em there is a unique HF state by design}:  
with $p$-filled CF Landau levels.

In the CF space there is room for another canonical pair besides $\bR_e$ 

\beq
\bR^v= \br +{l^2 \over c(1+c)}\hat{\bz}\times \bPi^*
\eeq

We call $\bR_v$ the guiding center coordinate of the double vortex
since it has the same charge $-2 \nu=-c^2$, as can be seen by the
commutator

\beq
 \left[ \bR^{v}_{x}, \bR^{v}_{y}\right]= i{l^2\over c^2}.
\eeq
Finally the two conjugate pairs  commute:
   \beqr
  \left[ \bR^{e}_{}, \bR^{v}_{}\right] &=& 0.
   \eeqr

Consider the densities formed by $\bR_v$:
\beqr
   \barchi &=& \sum_j \exp \left[ -i \bq \cdot \bR^{v}_{j}\right].
   \eeqr They   obey  
 \beq \left[ \bar{\chi}_{}(\bq ),
  \bar{\chi}_{}(\bq' )\right]= -2 i \sin \left[ {
    l^2\over 2c^2 }{\bq \times \bq' }\right] \bar{\chi}_{}(\bq +\bq') .\label{gmp2} \eeq 
    and commute with $\baro$:
    \beq
    \left[ \baro , \bar{\chi}_{}(\bq' ) \right]\equiv 0
    \eeq
  
We see that $\bar{H}$ commutes with a huge family of  operators $\barchi$ 
\beq
\left[ \bar{H}, \barchi \right]=0
\eeq

that do not enter $\bar{H}$ and close under commutation. The Appendix
shows that the $\barchi$ are the constraints that pay for the
magnetoplasmon oscillators that were introduced \`{a} la
Bohm-Pines\cite{Bohm-Pines} and decoupled.

If one wanted to skip the intermediate steps one could simply begin
with Eqns. \ref{Picom} - \ref{projH} which pose the LLL problem of
electrons in the larger space of the CF, {\em preserving the algebra of
  $\baro$ and $\bar{H}$}. The physical sector is defined by $\barchi
\simeq 0$ where $\simeq$ means ``when $\barchi$ appears within
correlation functions''.

While a HF state exists in the CF space, the result of naive HF
calculations is a mixed bag. On the plus side, one sees the
$K$-invariance of Haldane in the HF spectrum at $\nu=\half$. On the
minus side, $\bar{\rho}(\bq =0)$, the charge associated with $\baro$
seems to be $e$ and not $e^*= e (1-c^2)$, there is no evidence of the
dipole at $\nu = \half$, and the structure factor is $S(q)\simeq q^2$
and not $q^4$, as required by Kohn's theorem.  However, these features
can be recovered upon imposing the constraints via a conserving
approximation \`{a} la Kadanoff and Baym,\cite{Kadanoff-Baym} which
enforces $\barchi \simeq 0$.  When Read\cite{Read-nu=1} carried out
this approximation for $\nu =1$ bosons (also a CF Fermi liquid) he
found $S(q)\simeq q^3 \log q$ and the over-damped mode of HLR.
Murthy\cite{conserving-us} found $S(q) \simeq q^4$ for Jain fractions
using the conserving approximation.

\section{ The Preferred density$\bar{\rho}_p(\bq)$.}

At this point we proposed a short-cut to some of these results
obtained in the conserving approximation. We argued that in an exact
theory which obeyed the constraint, we could replace $\baro$ by $\baro
- \alpha \barchi$ for any value of $\alpha$. While all values of
$\alpha$ were equal in the exact theory, the following one stood out
as the {\em preferred density} in the HF calculation:

\beq \bar{\rho}_p= \baro - c^2 \barchi \eeq
 
With this choice, one found, as $q \to 0$, the charge $(1-c^2)e =
e^*$, $S(q)\simeq q^4$ and the correct dipole moment:

\beq \bar{\rho}_p(\bq)=\sum_j e^{-i \bq \cdot \br_j}\left((1-c^2)-il^2
\bq \times \bPi^{*}_{j}+ \ldots\right) \label{rhoexp}\eeq 

At $\nu = \half$, $c=1$, the fermion becomes neutral, $\bPi^{*}_{j}
\to \bp$ and we regain the dipole moment by Read's wave function
analysis.  In this approach using

\beq\bar{H}_p=\half 
\sum_{\bq} \bar{\rho}_p (\bq) v_{ee}(q) \bar{\rho}_p(-\bq) \eeq 

one gets many results (at small $\bq$) of the conserving approximation
{\em at tree-level}. On the other hand, $\barop$ does not obey the GMP algebra except at small $\bq$, and there is no 
systematic way of deriving the overdamped mode of HLR\cite{HLR}.

We will see shortly that this {\em ad hoc} recipe becomes a legitimate option in the algebraic approach {\em but only at $\nu =\half$} .

Going forward, it should be borne in mind that {\em when we work with
  $\bar{\rho}_p$ and $\bar{H}_p$, there are no commuting   constraints to select  states  CF space corresponding to electrons in the LLL .} Instead the role
of $\barchi$ is to represent the charge density of the
double-vortex. In the spirit of a low-energy effective description, we
hope, motivated by Jain\cite{Jain}, that despite the larger CF Hilbert
space, low energy properties of the FQH states in the LLL will be
correctly reproduced.
 
We addressed several quantitative questions\cite{us-preferred} using
$\bar{H}_p$ and found it to be a 10-20\% theory for gaps,
polarizations etc., as long as there was an infrared cut off in the
form of a gap or temperature.

\section{The CF-Fermi sea following Jain}

Let us focus on finding a formalism which exhibits the PH symmetry at $\nh$. This is a special point in other ways as well. Here the electron and vortex have exactly opposite charges,  the CF is electrically neutral and sees no magnetic field.  But even more special is the following: {\em 
the preferred density $\barop$ itself obeys the GMP algebra}:
\beqr
& &\left[\baro - \barchi , \bar{\rho}(\bq')- \bar{\chi}(\bq')\right]= \left[\baro , \bar{\rho}(\bq')\right]+\left[\bar{\chi}(\bq) , \bar{\chi}(\bq')\right]\nonumber\cr
&=& 2 i \sin \left[ {
    l^2\over 2 }{\bq \times \bq' }\right] \left(\bar{\rho}_{}(\bq +\bq')-\bar{\chi}_{}(\bq +\bq')\right).
\eeqr

Thus if we follow the algebraic route, {\em at $\nu = \half$, $\barop$
  is another candidate besides $\baro$ that satisfies the GMP
  algebra. } Thus the {\em ad hoc} introduction of $\barop$ as a short
cut to the results of the conserving approximation now becomes a
legitimate alternative to $\baro$.

In other words there are two ways to obtain a realization of the GMP algebra. The first is to find the electron guiding center coordinate in the CF space and then to exponentiate it, as in Eqns. \ref{Re} and \ref{expRe}. The other is to directly go for the preferred densities densities $ \bar{\rho}_p= \baro -  \barchi$ which are not exponentials of anything simple.

Thus, unlike $\baro$, which evolved adiabatically from the CS formulation,
the use of $\barop$ represents a leap   based entirely on algebraic
considerations. There is no reason to believe it has to be realized in
the space of the one-component fermion, or that if it is realized in
another space, that representation has any relevance to the original
LLL problem.

To begin with,  let us assume that $\barop$ lives in the space of the one-component fermion and see what happens.

First consider the  anti-unitary time-reversal operation  in this
representation. It is easy to see in first quantization: as $\bPi^*
\to \bp$ and $c=1$ at $\nu =\half$,

     \beqr
     \bR_e &=& \br -{l^2 \over 1+c}\hat{\bz}\times \bPi^*= \br -{l^2 \over 2}\hat{\bz}\times \bp \label{28}\cr
   \bR_v&=& \br +{l^2 \over c(1+c)}\hat{\bz}\times \bPi^*=\br +{l^2 \over 2}\hat{\bz}\times \bp \label{29}
   \eeqr
Note that 
\beqr
 {{\bR_e}+ \bR_v \over 2}&=&\br\ \mbox{(CF is midway between $e$  and $v$)} \label{midway}\cr
{{\bR_e}- \bR_v  }&=&-{l^2 }\hat{\bz}\times \bp \ \ \ \ \mbox{(CF dipole moment)}\label{cfdipole}
\eeqr

   Under {\em time-reversal} ${\cal T}$, we see that 
   \beqr
   {\cal T}: & & \bR_e \leftrightarrow \bR_v\cr
   {\cal T}: & & \bar{\rho}(\bq) \leftrightarrow \bar{\chi}(-\bq)\cr
   {\cal T}: & & \bar{\rho}_p(\bq) \to - \bar{\rho}_p(-\bq).
   \eeqr
  The last equation informs us that ${\cal T}$ has effected the PH transformation  on the electronic charge. 
The Hamiltonian being bilinear in $\bar{\rho}_p$
remains invariant. 

In second quantization where 
   \beqr
   \baro&=&\sum_{\bk}d^{\dag}_{\bk - \bq}e^{-\frac{il^2}{2}\bq \times \bk}d_{\bk}\cr
   \barchi&=&\sum_{\bk}d^{\dag}_{\bk - \bq}e^{\frac{il^2}{2}\bq \times \bk}d_{\bk}.
   \eeqr
   and the action of ${\cal T}$ is 
   \beqr
   {\cal T}d^{\dag}_{\bk} {\cal T}^{-1}&=& d^{\dag}_{-\bk}\cr
   {\cal T}d^{}_{\bk} {\cal T}^{-1}&=& d^{}_{-\bk}\cr
   {\cal T}i {\cal T}^{-1}&=& -i,
   \eeqr
   
   we have 
   \beqr
   {\cal T}\baro {\cal T}^{-1}&=& \sum_{\bk}d^{\dag}_{-\bk - \bq}e^{\frac{il^2}{2}\bq \times \bk}d_{-\bk}\cr
   &=& \sum_{\bk}d^{\dag}_{\bk - \bq}e^{-\frac{il^2}{2}\bq \times \bk}d_{\bk}\cr
   &=& \bar{\chi}(- \bq)
   \eeqr
   
   and likewise 
   \beq
   {\cal T}\barchi {\cal T}^{-1}=\bar{\rho}(- \bq).
   \eeq
Consequently 
   \beqr
   {\cal T}\bar{\rho}_p(\bq ) {\cal T}^{-1}&=&{\cal T}(\baro -\barchi) {\cal T}^{-1}\cr
   &=&-\bar{\rho}_p(-\bq ).
   \eeqr

So this symmetry reverses the sign of the (preferred) physical charge density
(represented now by $\bar{\rho}_p$), making  it  appropriate to call it
charge-conjugation.  This is reminiscent of  Son's approach\cite{Son}, but 
there is a crucial difference and we are grateful to both Son and
Senthil for emphasizing this: The anti-unitary operation ${\cal T}$ we
have proposed obeys

\beq
{\cal T}^2 = +1
\eeq

whereas the PH symmetry of the electronic LLL problem obeys\cite{levin-son}

\beq
{\cal T}^2 = -1.
\eeq

Consequently this model cannot be a representation of the electronic
$\nu =\half$ LLL problem. And yet it is a model in which there is a
CF-Fermi surface, which manifestly displays the GMP algebra of the
charge density, and the dipolar picture.  If it is not the $\nu =
\half$ problem of electrons, what is it? There is one obvious choice:
the $\nu=1$ hard-core boson problem analyzed by
Read\cite{Read-nu=1}. Indeed if we switch from the hard-core bosons to
CFs by attaching one unit of statistical flux and followed our
Bohm-Pines approach we would get the same expressions for $\baro$ and
$\barchi$ at small $q$. If we extended to small-q results to all $q$
to satisfy the GMP algebra, we would get exactly the charge density
and constraint algebra that Read\cite{Read-nu=1} obtained for the
boson problem. The dipole here is also made of charge $\pm 1$
objects. It can be quantized as a single-component fermion given the
absence of extra phase factors from the boson-vortex bound state.

If our $\bar{H}_p$ indeed describes the $\nu =1$ boson problem, it
suggests that there is an emergent charge-conjugation symmetry of the
$\nu=1$ boson problem as well. In addition we have a concrete
representation of the electronic charge density in $\barop$, which
permits us to do many of the detailed calculations of response
functions, even at $T>0$.

\section{Hamiltonian formulation of Son's Dirac CFs at $\nu=\half$}

To describe the electronic $\nh$ problem, we need ${\cal
  T}^2=-1$. This is impossible in the space of a single-component
fermion. Having seen that the representation of $\barop$ need not be
continuously connected to the primordial problem, we can seek other
options. We confess we could not have made any headway till we turned
to the very surprising option Son\cite{Son} provides us, of a Dirac fermion.
This option is buttressed by Wang and Senthil\cite{Wang-Senthil} who
give us nice a physical picture of why this is so: one of the double
zeros must lie on the electron (by the Pauli principle) turning it
into a charge $\half$ semion. The remaining vortex is a charge
$-\half$ anti-semion. The pair is quantized as a spinor (as shown in
the Appendix of WS\cite{Wang-Senthil}) that appears in Son's Dirac
equation. Given this internal structure the phase of $\pi$ due to
circumnavigating the Fermi circle follows.

One may feel that where we place the two vortices (none on the electron or
just one on the electron which makes it a semion) is a short distance
feature, that  both descriptions have the same long distance features: a net
charge of zero and the same dipole moment $\simeq \hat{\bz} \times \bp
l^2$ which appears at $\nu =\half$ fermions and $\nu =1$
bosons. However, the difference in
internal structure leads to a profound difference in the  topology of the Fermi
surface, one with a Berry phase and one without. 

Let us now implement our algebraic approach starting with a Dirac
fermion, {\em which will be our composite fermion.} The number of
these CFs, which so far equaled the number of electrons, is also equal
to half the number of flux quanta penetrating the sample precisely at
$\nh$, exactly as in Son's construction\cite{Son}. {\em Thus $\nu
  =\half$ is the confluence of two approaches, the one we have always
  used, in which $N_{CF}= N_{e}$ and Son's in which $N_{CF}= \half
  N_{\phi}$.} We will see that if we are to go to $\nu\ne \half$ we
must follow Son's assignment.

In 2+1 dimensions the
noninteracting two-component Dirac equation is

\beq
i\hbar \partial_t \psi=\sigmab\cdot (\bp -\ba) \ \psi
\eeq
where $\ba$ is an external gauge field. Let us initially set it $\ba =0$. 

As usual, there are positive and negative energy solutions. Paying no
attention to filling the Dirac sea, we can expand $\psi$ in real space
as

\beq
\psi^T (\br ) = \sum_{\bk} e^{i \bk \cdot \br} \left[ {c_{\bk}\over \sqrt{2}}  (1, e^{i\theta_{\bk}})+{d_{\bk}\over \sqrt{2}} (1, -e^{-i\theta_{\bk}})\right].
\eeq
where $e^{i\theta_{\bk}}={k_x+ik_y\over|k|}$.

The projected electron density is of course not the density of the
Dirac CF, just like it was not in the previous case of the
non-relativistic field. It is determined by the GMP algebra. Let us start 
with the same expressions for $\bR_e$ and $\bR_v$ as  in
Eqns. (\ref{28}) and (\ref{29}) adapted at $\nu = \half$ (to fermions in zero field), 

  \beqr
     \bR_e &=&  \br -{l^2 \over 2}\hat{\bz}\times \bp \label{288}\cr
   \bR_v&=& \br +{l^2 \over 2}\hat{\bz}\times \bp \label{298}
   \eeqr 
Note that under time-reversal 
\beq
{\cal T}: \ \ \bR_e \leftrightarrow \bR_v.
\eeq

Next we define the electron density and
vortex density operators in the Hilbert space of the Dirac CF by taking matrix elements of $e^{-i\bq\cdot\bR_e}$ and $e^{-i\bq\cdot\bR_v}$ between momentum states and  obtain


\beqr
\baro \!\! &=&\!\! \half \sum_{\bk} e^{-i\half l^2 \bq \times \bk}\left[ \left[ c^{\dag}_{\bk -\bq}c_{\bk}+
d^{\dag}_{\bk -\bq}d_{\bk} \right]\!(1 \!+ \!e^{i(\theta_{\bk} - \theta_{\bk -\bq})})\right.\nonumber\cr
&+& \!\! \left.\left[ c^{\dag}_{\bk -\bq}d_{\bk}+
d^{\dag}_{\bk -\bq}c_{\bk} \right](1 - e^{i(\theta_{\bk} - \theta_{\bk -\bq})})\right]
\label{barrhodirac}\eeqr

and
\beqr
\barchi \!\!&=&\!\! \half \sum_{\bk} e^{
i\half l^2 \bq \times \bk}\left[ \left[ c^{\dag}_{\bk -\bq}c_{\bk}+
d^{\dag}_{\bk -\bq}d_{\bk} \right]\!(1 \!+\! e^{i(\theta_{\bk} - \theta_{\bk -\bq})})\right.\nonumber\cr
&+& \!\! \left.\left[ c^{\dag}_{\bk -\bq}d_{\bk}+
d^{\dag}_{\bk -\bq}c_{\bk} \right](1 \!- \!e^{i(\theta_{\bk} - \theta_{\bk -\bq})})\right]\!\! 
\label{barchidirac}\eeqr

One may verify that  $\baro$ and ${\bar
  \chi}(\bq )$ obey the same algebra as before as before since $\bR_e$ and $\bR_v$ do. So this is yet another algebraically faithful
representation of the LLL. Once again we will use our preferred
density $\bar{\rho}_p= \bar{\rho}- \bar{\chi}$, ignoring constraints in the spirit of obtaining a low-energy
theory that has all the symmetries of the original. It is  $\bar{\rho}_p$ that allows us to display the PH transformation  as follows. 

The  PH transformation  of electrons is once again implemented as
time-reversal on the CF.  However the action of ${\cal T}$ 
naturally follows from the Dirac nature of the CF

\beqr
{\cal T}c_{\bk}{\cal T}^{-1}&=& e^{i \theta_{\bk}}c_{-\bk}\cr
{\cal T}c_{\bk}^{\dag}{\cal T}^{-1}&=& e^{-i \theta_{\bk}}c^{\dag}_{-\bk}\cr
{\cal T}d_{\bk}{\cal T}^{-1}&=& -e^{i \theta_{\bk}}d_{-\bk}\cr
{\cal T}d_{\bk}^{\dag}{\cal T}^{-1}&=& -e^{-i \theta_{\bk}}d^{\dag}_{-\bk}
\eeqr
Using $\theta_{\bk}- \theta_{-\bk}=\pi$, one sees that ${\cal T}^2=-1$. Next, 
one may verify that 
\beqr
{\cal T}\baro{\cal T}^{-1}&=&\bar{\chi}(-\bq)\cr
{\cal T}\barchi{\cal T}^{-1}&=&\bar{\rho}(-\bq)\cr
{\cal T}\bar{\rho}_p(\bq) i{\cal T}^{-1}&=& - \bar{\rho}_p(-\bq)
\eeqr
(This is most easily seen in first-quantization by considering the action of ${\cal T}$ on $e^{-i \bq \cdot \bR_e}$ and $e^{-i \bq \cdot \bR_v}$: for  $i \bq \to -i \bq$ and $\bR_e \leftrightarrow \bR_v$.)

Since ${\cal T}$ reverses the sign of the electronic charge $\bar{\rho}_p$  it is
appropriate to call it a PH transformation. The Hamiltonian built out of the preferred density 
\beqr
\bar{H}_p&=&\half\sum\limits_{\bq}v_{ee}(q)\bar{\rho}_p(\bq)\bar{\rho}_p(-\bq)
\eeqr
is symmetric under ${\cal T}$. 

On the other hand the Hamiltonian built out of $\bar{\rho}$
\beq 
\bar{H}=\half\sum\limits_{\bq} v_{ee}(q){\bar{\rho}}(\bq){\bar{\rho}}(-\bq)
\eeq
which does not display PH symmetry but has a huge symmetry group generated by $\bar{\chi}$
is the only way to get the overdamped mode in a conserving calculation.  

(Filling the Dirac sea will lead to the replacements $d_{\bk}\to
b^{\dagger}_{-\bk},\ \ d^{\dagger}_{\bk}\to b_{-\bk}$, where the
$b,\ b^{\dagger}$ are now hole destruction and creation operators for
negative energy states in the filled Dirac sea. If one is
interested in nonzero filling above the Fermi point in the Dirac
problem, the $d$ or $b$ operators are high energy operators and can be
set to zero to obtain the low-energy physics. However, the density, if
projected by setting $d,d^{\dagger}\to 0$, will not obey the GMP
algebra exactly, but only at small $q$.)

It should be noted that our CF differs from Son's\cite{Son} in one regard. 
The projected physical charge 
density of the {\em correlated} electrons  is directly given in terms of our CFs but in
Son's picture, the physical number density of electrons (measured from half-filling)  is the curl of an emergent, minimally coupled  gauge
field.

Next we turn to the nature of the ground state of the interacting Hamiltonian we have proposed. 

\section{Hartree-Fock nature of Son's ground state}

So far, all we have shown is that we can realize the GMP algebra in a
Hilbert space of Dirac fermions. However, unlike the effective theory
proposed by Son\cite{Son}, which already comes with a ``kinetic'' term for the
Dirac fermions, our two proposed interacting Hamiltonians

\beqr
\bar{H}=&\half\sum\limits_{\bq} v_{ee}(q){\bar{\rho}}(\bq){\bar{\rho}}(-\bq)\cr
\bar{H}_p=&\half\sum\limits_{\bq}v_{ee}(q)\bar{\rho}_p(\bq)\bar{\rho}_p(-\bq)
\eeqr

have no such kinetic terms. So one may ask in what sense one can make
a correspondence between Son's proposed ground state (all negative
energy states filled, and positive energy states filled to some $\mu$
which guarantees the correct number of CFs).

The answer is that Son's ground state is a Hartree-Fock state of both
of our interacting Hamiltonians. To see this we characterize Son's
ground state in terms of the expectation values of the $c$ and $d$ operators defined in the previous section:

\beqr
\langle d^{\dagger}_{\bk}d_{\bk'}\rangle = &\delta_{\bk\bk'}\ \ \ \forall \bk\cr
\langle c^{\dagger}_{\bk}c_{\bk'}\rangle = &\delta_{\bk\bk'} N_{Fc}(\bk)\cr
\langle d^{\dagger}_{\bk}c_{\bk'}\rangle = &\langle c^{\dagger}_{\bk}d_{\bk'}\rangle = 0
\eeqr

where $N_{Fc}(\bk)=\Theta(k_F-k)$. 
One now writes the interacting Hamiltonian and reduces it to a
one-body (HF) Hamiltonian by taking all possible expectation
values. One can see by inspection that since translation symmetry is
preserved by the ground state, the HF Hamiltonian must be of the form

\beqr
H_{HF}=&\sum\limits_{\bk} \big(\epsilon_c(\bk)c^{\dagger}_{\bk}c_{\bk}+\epsilon_d(\bk)d^{\dagger}_{\bk}d_{\bk}\nonumber \cr
&+\gamma(\bk)c^{\dagger}_{\bk}d_{\bk}+\gamma^*(\bk)d^{\dagger}_{\bk}c_{\bk}\big)
\eeqr

Each of the coefficients $\epsilon_{c,d}(\bk),\ \gamma(\bk)$ are sums
over $\bq$. If the coefficients $\gamma(\bk)$ are not zero one
generates correlations between $c$ and $d$, and the ground state
proposed by Son will fail to be a HF state of our interacting
Hamiltonian. So the verification that Son's ground state is a HF state of our Hamiltonian reduces to verifying that $\gamma(\bk)=0$. A straightforward calculation shows that in  the case of  $\bar{H}$

\beqr
\gamma(\bk)=&-\frac{i}{4}\sum\limits_{\bq} v_{ee}(q)\big(\sin(\theta_{\bk}-\theta_{\bk-\bq})[1-N_{Fc}(\bk-\bq)]\nonumber\\
&+\sin(\theta_{\bk}-\theta_{\bk+\bq})[1-N_{Fc}(\bk+\bq)]\big)
\eeqr

Now, we note that for every $\bk,\ \bq$, there is a $\bq^*$ which is
the vector $\bq$ reflected about $\bk$. All the terms in the
expression for $\gamma$ are even under the change $\bq\to\bq^*$,
except for the prefactor $\sin(\theta_{\bk}-\theta_{\bk\pm\bq})$ which
changes sign. The sum is thus zero by symmetry for any rotationally
invariant $v_{ee}(q)$. 

A similar arguments applies for the case $\bar{H}_p$ in which we use the preferred density $ \bar{\rho}_p= \baro -  \barchi$, despite additional phase factors.

Examining the HF Hamiltonian in more detail reveals that for any
$\bk$, $\epsilon_d(\bk)\le\epsilon_c(\bk)$. Equality is achieved only
for $\bk=0$. These are also features of the noninteracting ground
state of Son.

So we have established that Son's ground state is a HF ground state of
our interacting Hamiltonians $\bar{H}$ and $\bar{H}_p$. We plan to use
$\bar{H}$, which commutes with $\bar{\chi}$, as a starting point for a
conserving calculation of response functions in the near future. As
for $\bar{H}_p$, there are no obvious constraints that commute with
it. However, in addition to manifestly displaying the PH symmetry, it
captures many of the low energy properties of the CF at the level of
naive Hartree-Fock.

\section{Hamiltionian formulation of Son's CFs away from $\nu=\half$}

The construction we carried out at $\nu=\half$ can be extended in a
natural way away from $\half$. But this requires us to change our
strategy. Until recently we were of the view that $\barop$ existed as
an alternative to $\barop$ only at $\nu = \half$. This is in fact true
if we insist on $N_{CF}= N_e$. However a new path opens up if we
switch to $N_{CF}= \half N_{\phi}$. Here are the details.

Recall again that the number of Son's
CFs is
 \beq N_{CF}={N_{\phi}\over2}={eBA\over4\pi}.
\eeq
 The CF couples to electronic charge via a gauge
potential whose curl is the physical charge density. The effective
number of flux quanta seen by the CFs is
 \beq
 N_{\phi,CF}=N_{\phi}-2N_e,
\eeq

where $N_e$ is the number of electrons. It follows that the effective
magnetic field seen by the CFs is 
\beqr B_{CF}&=& B(1-2\nu)\equiv B\ \theta
\eeqr
where 
\beq
\theta = 1 - 2\nu.
\eeq

 Introducing coordinates $\br^{CF}$ and velocity operators $\bPi^{CF}$ we demand

  \beq
\left[\bPi^{CF}_x(\theta ) ,\bPi^{CF}_y (\theta ) \right]={i(1-2\nu)\over l^2}\equiv {i\theta\over l^2}\label{PICOMM}
\eeq

where 
\beqr
\bPi^{CF}(\theta) &=&\bp - \ba (\theta ).
\eeqr

If we do a PH transformation {\em on electrons}, we want  $ \theta  \to -\theta $ and the gauge field (which represents the electron density away from $\nh$) to reverse its sign:  

\beqr
 \theta &\to& -\theta \cr
\ba (\theta) & \to& -\ba (\theta ) =\ba (-\theta).
\eeqr
These changes are implemented in the CF world by ${\cal T}$. First of all 
\beqr
{\cal T}\bPi^{CF}(\theta ){\cal T}^{-1} &=& {\cal T}(\bp - \ba (\theta )){\cal T}^{-1}\cr
&=& (-\bp - \ba (\theta ))\cr
&=& - (\bp + \ba (\theta) )\cr
&=& - (\bp - \ba (-\theta) )\cr
&=& - \bPi^{CF}(-\theta).
\eeqr
If we now conjugate Eqn. \ref{PICOMM} we find, using ${\cal T} i {\cal T}^{-1}= -i$, 
\beq
[\bPi^{CF}_x(-\theta ) ,\bPi^{CF}_y (-\theta) ]= {-i\theta\over l^2}
\eeq
 as desired. 

At $\theta \ne 0$, we define two sets of conjugate coordinates $\bR^e$ and $\bR^v$
as follows

\beqr \bR^e(\theta )=&(1+{\theta\over 4})\br^{CF}-\half
l^2\zh\times\bPi^{CF}(\theta )\cr \bR^v (\theta) =&(1-{\theta\over4})\br^{CF}+\half
l^2\zh\times\bPi^{CF} (\theta ).\eeqr

It can easily be checked that {\em for all $\theta$}, 
\beq
[R^e_x,R^e_y]=-il^2,\ \ \ \  [\bR^e,\bR^v]=0, \ \ \ \ 
[R^v_x,R^v_y]={il^2}. 
\eeq
There are several pleasing features of
these sets of coordinates. Firstly, under time-reversal in the Dirac
world, since  ${\cal T}\bPi (\theta ) {\cal T}^{-1}=-\bPi (-\theta)$
we find
\beqr
{\cal T}\bR^e(\theta ) {\cal T}^{-1}&=&(1+{\theta\over 4})\br^{CF}+\half
l^2\zh\times\bPi^{CF}(-\theta)\cr &=& \bR^v (-\theta)\label{etov}\cr
{\cal T}\bR^v(\theta ){\cal T}^{-1}&=&(1-{\theta\over 4})\br^{CF}-\half
l^2\zh\times\bPi^{CF}(-\theta)\cr &=& \bR^e (-\theta).\label{vtoe}
\eeqr
Secondly, the position coordinate of the
Dirac CF is still the average of $\bR^e$ and $\bR^v$, as at
$\nu=\half$.

Finally  we define $\bar{\rho}$ and $\bar{\chi}$  in the Hilbert space of
the Dirac CFs exactly as before, by taking the matrix elements of $e^{-i\bq\cdot\bR_e}$ and $e^{-i\bq\cdot\bR_v}$ between momentum states of the Dirac
fermion.


Because the commutation relations of $\bR^e$ and $\bR^v$ are identical
to those at $\nu=\half$, we can once again choose to represent the
physical charge density in two distinct ways, either as ${\bar\rho}$
or as ${\bar\rho}_p={\bar\rho}-{\bar\chi}$.

If we choose to represent the Hamiltonian in terms of ${\bar\rho}$ it
will commute with the set of ${\bar\chi}(\bq)$, and will thus be
amenable to a conserving approximation\cite{Kadanoff-Baym}. Of course,
the physical charge $e^*$ of the quasiparticles and the PH mapping
will not be manifest.

If, on the other hand, we choose to represent the physical charge
density as ${\bar\rho}_p$ then the PH transformation (implemented by
${\cal T}$ with ${\cal T}^2=-1$) can be explicitly realized as
follows.  Given Eqn. \ref{etov} and \ref{vtoe}: 

\beqr {\cal
  T}{\bar\rho}(\bq , \theta){\cal T}^{-1}&=&{\bar\chi}(-\bq ,
-\theta)\cr {\cal T}{\bar\chi}(\bq , \theta){\cal
  T}^{-1}&=&{\bar\rho}(-\bq , -\theta)\cr {\cal T}{\bar\rho}_p(\bq ,
\theta){\cal T}^{-1}&=&-{\bar\rho}_p(-\bq , -\theta).  \eeqr 

The
Hamiltonian, quadratic in $\barop$ will respond as follows: \beq
\bar{H}_p(\theta) \to \bar{H}_p(-\theta).  \eeq

An important point to note: the {\em ad hoc} combination $\bar{\rho}_p
= \barop - c^2 \barchi$ stood for a particle of charge $e^*=e(1-2\nu )
$, with $e^*=0$ only at $\nu = \half$. In the present approach
mirroring Son's, $\bar{\rho}_p$ always describes a neutral particle
($\bar{\rho}_p (\bq =0)=0$). This is actually too much of a good
thing, since the actual charge of the quasiparticle at long
wavelengths for $\nnh$ should be $e^*\ne0$. This is a problem we hope
to resolve in future work.

Thus, we have been able to find a representation of the GMP algebra in
terms of the neutral Dirac CFs of Son\cite{Son} for all $\nu$.
Presumably, this could be the starting point for calculations of gaps,
response functions, etc, as in our previous work on Jain's CFs.

\section{Summary}

This paper explores the Son's\cite{Son} recent solution for displaying the PH
{\em transformation} of electrons in the LLL and the PH {\em symmetry}
of the $\nh$ problem, within the framework of our Hamiltonian
formalism.  In our approach we map the algebraic problem of the LLL
projected charge $\baro$ (which obeys the GMP algebra) and the
projected Hamiltonian $\bar{H}( \baro )$, from the electronic space
(plagued with ground state degeneracy) to a different space which
permits a unique HF ground state. When realized in the CF space of a
single-component fermion which saw the weaker field mandated by Jain\cite{Jain},
we also found a closed algebra of constraints $\barchi$ that commuted
with $\bar{H}$, delineated the physical sector, and formed the GMP
like algebra of an object with the charge $-2\nu$ of the double
vortex. These results could be derived at small $q$, as indicated in
the Appendix.

Recently we realized that {\em at and only at} $\nh$, a preferred
charge density $\barop = \baro - \barchi$ also obeyed the GMP algebra
and could equally well represent the projected, correlated electron
density. The role of this isolated second option escaped us until
recently, as did the importance of the PH transformation of the LLL.
We now see that it allows us to realize the PH transformation as an
anti-unitary operator ${\cal T}$ with ${\cal T}^2=+1$ in the space of
a one-component fermion. Since one wants ${\cal T}^2=-1$ for
electrons\cite{levin-son,gang-of-six}, we conjecture this describes an emergent
symmetry of the $\nu=1$ boson problem studied by Read\cite{Read-nu=1}.

Following Son\cite{Son} and the work of Wang and
Senthil\cite{Wang-Senthil} we then cast the algebraic formulation in
the space of Dirac fermion.  By following Son's approach of equating
the number of CFs to half the number of flux quanta (and not the
number of electrons) we were able to extend the operator approach to
all $\nu$. In this version the CF is always neutral. The commutation
relations for $\baro$ and $\barchi$ are the same at $\nh$. As always,
we have two options. One is to use $\baro$ as the electronic charge
density and $\barchi$ as the algebra of constraints that specifies the
physical LLL sector.  The constraints are then to be enforced in a
conserving approximation\cite{Kadanoff-Baym}, which would yield the
overdamped mode at $\nh$.  The other option is to use $\barop$, in
which the PH transformation properties are transparent. However,
unlike the $\barop$ of the one-component (Jain) CFs, the quasiparticle
charge does not come out correctly at tree level. Perhaps there is an
even better representation in which all the algebraic and symmetry
properties of the CF are manifest. 

There are a number of future directions we would like to pursue. The
first is to carry out a conserving calculation at $\nh$ in the new
formulation in terms of Dirac CFs. We have already established the
first necessary step, that Son's ground state is a HF ground state of
our interacting Hamiltonian.  The structure factor should vanish as
$q^4$ to be in compliance with Kohn's theorem. We should also recover
the overdamped mode of HLR\cite{HLR}, and in the presence of disorder
we should be able to see the suppression of
backscattering\cite{gang-of-six}. Presumably, we should be able to
extend this kind of treatment to $\nu$ away from half as well. At the
moment, we have a realization of the GMP algebra away from $\half$
that does show the mapping from $\nu\to1-\nu$. However, the density
operator does not have the correct quasiparticle charge. We would like
to find a representation in which all the algebraic, symmetry, and
physical properties of the quasiparticles of the problem are
transparently visible.

We are grateful to Son for extended discussions at the Aspen Center
for Physics.  We thank Senthil for sharing with us his work with Wang
and the physical picture in terms of semions.  His input, through
extensive email and phone communications, was crucial to our
understanding of the central issues. RS thanks Ashvin Viswanath for
conversations.  We thank the Aspen Center for Physics, where this work
was conceived and completed, and which is supported by National
Science Foundation grant PHY-1066293. Murthy also acknowledges partial
support from the NSF via DMR-1306897 (GM) and from the US-Israel
Binational Science Foundation via grant no. 2012120.

\section{Appendix}
In our earliest work we began with the CS theory and adapted the
 strategy of Bohm and Pines. Collective charge degrees of freedom were
 represented by magnetoplasmon oscillators $A(\bq)$ of cyclotron
 energy $\omega_c$ and to pay for them some constraints $\chi (\bq)$
 were imposed. At the end the fermions and oscillators were decoupled
 in the small $q$ limit to yield the following results for Jain
 fractions: 
 \beqr H_{osc}&=& \sum_{\bq}A^{\dag}(\bq)A(\bq)\omega_c\cr \bj_e (\bq)&=&
 \hat{\bq} (A(\bq)+A^{\dag}(\bq))\cr
  \rho_e(\bq )&=& q(A (\bq)+ A^{\dag}(\bq) )+
 \baro \ \ \mbox{where}\cr \baro &=& \sum_j\e^{-i\bq \cdot
   \br_j}\left(1- {il^{2}\over 1+c}\bq \times
 \bPi^{*}_{j}+\ldots \right)\ \ \ \mbox{}\cr \barchi &=& \sum_j\e^{-i\bq \cdot
   \br_j}\left(1+ {il^{2}\over c(1+c)}\bq \times
 \bPi^{*}_{j}+\ldots \right)\ \ \ \cr 0&=& \barchi |\mbox{physical state
 }\rangle \ \ \ \mbox{(constraint)} \eeqr where 
 \beq c^2 = 2 \nu ={2p
   \over 2p+1}, \eeq and $\bPi^{*}_{j}$ is the canonical momentum of CF
 number $j$ which experiences the right field to satisfy Jain's
 condition \beq \left[ \bPi^{*}_{x}, \bPi^{*}_{y}\right] = {i
   (1-c^2)\over l^2}\equiv {i\over l^{*2}}.  \eeq Notice that the current is
 carried only by the oscillators at every $\nu$. When the charge
 $\rho_e$ is coupled to an external scalar potential $\Phi(\bq)$, the
 resultant Hall current gives the correct $\sigma_{xy}$. Our CF,
 restricted the LLL makes no contribution since the current has no
 leading matrix elements within the LLL.
   
   Following this, it was conjectured by RS that the two terms in the expression for $\baro$ and $\barchi$ were the beginnings of  the following  exponentials  
   \beqr
   \baro &=& \sum_j \exp \left[ -i \bq \cdot \bR^{e}{j}\right]\cr
   \barchi &=& \sum_j \exp \left[ -i \bq \cdot \bR^{v}_{j}\right]\ \ \mbox{where}\cr
   \bR^e &=& \br -{l^2 \over 1+c}\hat{\bz}\times \bPi^*\cr
   \bR^v&=& \br +{l^2 \over c(1+c)}\hat{\bz}\times \bPi^*
   \eeqr
   In this ``all $\bq$'' formalism, the  coordinates $\bR^e$ and $\bR^v$ were named thus 
 because they have the following commutation relations:
   \beqr
   \left[ \bR^{e}_{x}, \bR^{e}_{y}\right] &=& -il^2\cr
   \left[ \bR^{v}_{x}, \bR^{v}_{y}\right] &=& i{l^2\over c^2}\cr 
   \left[ \bR^{e}_{}, \bR^{v}_{}\right] &=& 0.
   \eeqr
   We recognize the commutation rules of  $\bR^e$ as that of  the guiding center-coordinate of the electron and $\bR^v$ as describing the guiding center coordinate of the double vortex. This ensures that  the density corresponding to $\bR^e$, 
   \beq
   \bar{\rho}_{}(\bq )=\sum_je^{-i \bq \cdot \bR^{e}_{j}}
   \eeq
    obeys
 the GMP algebra. The density formed from the vortex coordinate obeys 
 \beq \left[ \bar{\chi}_{}(\bq ),
  \bar{\chi}_{}(\bq' )\right]= -2 i \sin \left[ {
    l^2\over 2c^2 }{\bq \times \bq' }\right] \bar{\chi}_{}(\bq +\bq') .\label{gmp} \eeq 
    and commutes with $\baro$:
    \beq
    \left[ \baro , \bar{\chi}_{}(\bq' ) \right]\equiv 0
    \eeq
and therefore with $\bar{H}(\baro )$. 

The careful reader will note that these results apply equally well to the $\nu =1$ boson problem after the fluxes are attached to the bosons.
  
\end{document}